\documentclass[11pt]{article}
\usepackage{amsmath,amssymb,color,epsfig,cite}
\usepackage{xcolor}
\usepackage{mathrsfs}

\usepackage{mathtools}

\usepackage{amsfonts}
\def\td{\tilde}
\newcommand{\hoch}[1]{$\, ^{#1}$}

\makeatletter
\@addtoreset{equation}{section}
\makeatother

\newcommand{\be}{\begin{equation}}
\newcommand{\ee}{\end{equation}}
\newcommand{\bea} {\begin{eqnarray}}
\newcommand{\eea}{\end{eqnarray}}
\newcommand{\nn}{\nonumber}

\def\ft#1#2{{\textstyle{\frac{\scriptstyle #1}{\scriptstyle #2} } }}
\def\fft#1#2{{\frac{#1}{#2}}}

\def\0{{\sst{(0)}}}
\def\1{{\sst{(1)}}}
\def\2{{\sst{(2)}}}
\def\3{{\sst{(3)}}}
\def\4{{\sst{(4)}}}
\def\5{{\sst{(5)}}}
\def\6{{\sst{(6)}}}
\def\7{{\sst{(7)}}}
\def\8{{\sst{(8)}}}
\def\sst#1{{\scriptscriptstyle #1}}

\def\del{{\partial}}

\def\cF{{{\cal F}}}

\def\wtd{\widetilde}

\def\ie{{\emph{i.e.~}}}

\def\ben{\begin{equation}}
\def\bea{\begin{eqnarray}}
\def\een{\end{equation}}
\def\eea{\end{eqnarray}}

\def\ft#1#2{{\textstyle{\frac{\scriptstyle #1}{\scriptstyle #2} } }}
\def\fft#1#2{{\frac{#1}{#2}}}

\begin{document}

\begin{flushright}
\hfill {UPR-1231-T\ \ \ \ CERN-TH-2022-112\ \ \ \ MI-HET-781}\\
\end{flushright}

\begin{center}

{\large {\bf Long-Range Forces Between Non-Identical Black Holes \\ With Non-BPS Extremal
Limits}}

\vspace{15pt}
{\large S. Cremonini$^{1}$, M. Cveti\v c$^{2,3,4,5}$, 
              C.N. Pope$^{6,7}$ and A. Saha$^{6}$}

\vspace{15pt}

{\hoch{1}}{\it Department of Physics, Lehigh University, Bethlehem, 
PA 18018, USA}

{\hoch{2}}{\it Department of Physics and Astronomy,
University of Pennsylvania, \\
Philadelphia, PA 19104, USA}

{\hoch{3}}{\it Department of Mathematics, University of Pennsylvania, Philadelphia, PA 19104, USA}

{\hoch{4}}{\it Center for Applied Mathematics and Theoretical Physics,\\
University of Maribor, SI2000 Maribor, Slovenia}

{\hoch{5}}{\it CERN Theory Department, CH-1211 Geneva, Switzerland}


\hoch{6}{\it George P. \& Cynthia Woods Mitchell  Institute
for Fundamental Physics and Astronomy,\\
Texas A\&M University, College Station, TX 77843, USA}

\hoch{7}{\it DAMTP, Centre for Mathematical Sciences,
 Cambridge University,\\  Wilberforce Road, Cambridge CB3 OWA, UK}

\vspace{10pt}

\end{center}




\begin{abstract}

 Motivated by recent studies of long-range forces beween identical black
holes, we extend these considerations by investigating the forces between
two non-identical black holes.  We focus on classes of theories where
charged black holes can have extremal limits that are not BPS.  These
theories, which live in arbitrary spacetime dimension, 
comprise
gravity coupled to $N$ 2-form field strengths and $(N-1)$ scalar
fields.  In the solutions we consider, each field strength carries
an electric charge.  The black hole solutions are governed by 
the $SL(N+1,R)$
Toda equations.  
In four dimensions the black-hole solutions 
in the $SL(3,R)$ example are equivalent to the ``Kaluza-Klein dyons.'' 
We find that any pair of such extremal
black holes that are not identical (up to overall scaling) 
repel one another.  We also show that there can exist pairs of
non-extremal, non-identical black holes which obey a zero-force condition.
Finally, we find indications of similar results in the higher 
examples, such as $SL(4,R)$.

\end{abstract}

{\scriptsize 
cremonini@lehigh.edu, cvetic@physics.upenn.edu,
pope@physics.tamu.edu, aritrasaha@physics.tamu.edu.}

\pagebreak

\tableofcontents
\addtocontents{toc}{\protect\setcounter{tocdepth}{2}}

\section{Introduction}

There has been a considerable interest recently in finding detailed ways to quantify the idea that gravity is the weakest force. Such attempts have led to a number of conjectures, 
with the Weak Gravity Conjecture (WGC) \cite{Arkani-Hamed:2006emk}  thus far on the strongest footing 
(see \cite{Harlow:2022gzl} for a comprehensive review). 
A natural way to examine the strength of gravity in a given theory is to study the force between well-separated particles and ask whether the gravitational attraction is indeed overwhelmed by the repulsive interactions in the theory.
This idea has been made more precise by the Repulsive Force Conjecture (RFC), which was originally stated in \cite{Arkani-Hamed:2006emk}  and describes the simple notion that long-range repulsive gauge forces (between identical charged particles) should be at least as strong as all long-range attractive forces. 
The RFC was reemphasized more recently in \cite{Palti:2017elp} and stronger versions were put forth by 
\cite{Heidenreich:2019zkl}. 
In its weakest form, the RFC argues that effective field theories (EFTs) consistent with quantum gravity must have a state which is ``self-repulsive," whether the state is a fundamental particle or a black hole.
When multiple charges are present, this requirement must then hold along each direction in charge space\footnote{The strongest form of the RFC \cite{Heidenreich:2019zkl} states that 
in every direction in charge space there should be a \emph{strongly self-repulsive multiparticle state}, \emph{i.e.} a multiparticle state where each constituent state repels every other state, including itself.}. 
While the WGC and the RFC are clearly related to each other, they are distinct in the presence of scalars \cite{Heidenreich:2019zkl}. Moreover, studies of black holes in theories with higher derivatives make the difference even more manifest \cite{Cremonini:2021upd}, and suggest that the 
RFC can \emph{not} be satisfied by the black hole spectrum alone -- unlike the WGC, which can \cite{Kats:2006xp}. 
Taking these considerations into account, it is valuable to 
better understand what else we can extract from the behavior of long-range forces, 
in order to eventually clarify to what extent the RFC is a useful criterion for constraining EFTs.

Within the context of the RFC, the interest thus far has been focused on long-range forces between two \emph{identical} copies of the same object. 
Two identical black holes at rest and asymptotically far from each other will generically attract, except at extremality, when the net force between them will vanish \cite{Heidenreich:2020upe}.
This is true even in theories with moduli, which mediate new long-range interactions and 
affect the 
balance of forces between the black holes\footnote{In theories with higher derivative corrections this is no longer true. The long-range force between extremal black holes receives corrections  and does not generically vanish \cite{Cremonini:2021upd}.}. 
In the simple case of a black hole of mass $M$ and electric charge $Q$ in theories in four dimensions with a single light scalar, the force between two identical copies is\footnote{We are neglecting  $\mathcal{O}\left(\frac{1}{r^3}\right)$
terms as well as velocity dependent forces.}
\begin{align}
\label{introforce}
    F(r) =   -  \frac{1}{4}\frac{M^2}{r^2} +  \frac{Q^2}{r^2} - \frac{\Sigma^2}{r^2} + \ldots
\end{align}
where $\Sigma$ denotes the \textit{scalar charge}  (to be defined precisely later). 
The contribution of the scalar field precisely offsets the gravitational and electrostatic forces between the black holes, ensuring that all interactions cancel.

Such a no-force condition, which here applies to any pair of identical, 
extremal, black holes, is also well-known from studies of 
supersymmetric BPS black holes in supergravity, where there exist multi-black hole solutions where the individual black holes sit in 
static equilibrium with zero force between them. 
However, in the case of extremal BPS black holes in supergravity, the no-force condition holds \emph{regardless of whether the black holes are identical or they instead carry different charges}.
This simple observation motivates us to examine more generally the behavior of long-range forces between two black holes that carry \emph{different} charges and 
are therefore not identical, working with configurations which are not BPS solutions. 
We are particularly interested in identifying any generic features of 
such interactions 
and under which conditions the force can vanish.
As we shall see, for non-identical black holes the behavior of the long range force is quite rich, and can still lead to zero force conditions, albeit in a different manner from the BPS cases arising in supergravity.  

A well-known example of a configuration with is not BPS is provided by the
so-called ``Kaluza-Klein dyon,'' which is a solution of 
four-dimensional ungauged ${\cal N}=8$ supergravity, in which a single
gauge field carries both electric and magnetic charge.  The
solution can be described by a consistent truncation of the 
full ${\cal N}=8$ theory that may instead be obtained as the Kaluza-Klein
reduction of pure five-dimensional Einstein gravity; hence the
nomenclature ``Kaluza-Klein dyon.''  This theory is described by
the four-dimensional Lagrangian
\bea
{\cal L}_4 &=& R -\ft12 (\del\phi)^2 -\ft14 e^{\sqrt3\phi}\, F^2\,.
\label{kkdyonlag}
\eea
For our purposes, it will be
more convenient to consider a somewhat different theory comprising gravity,
the dilaton and two distinct gauge fields rather than just a single 
gauge field.  Among other things, this has the advantage that we can
describe charged black holes with the feature of having an extremal but not
BPS limit in any arbitrary dimension $d$. The $d$-dimensional 
Lagrangian is given by \cite{luyang}
\bea
{\cal L} = R -\ft12 (\del\phi)^2 -\ft14 e^{a\phi}\, F_1^2 -
    \ft14 e^{-a\phi}\, F_2^2\,,\label{sl3rlag}
\eea
where 
\bea
a= \sqrt{\fft{6 (d-3)}{d-2}}\,.\label{acon}
\eea
The black hole solutions we shall consider have two independent electric
charges, one carried by each of the two field strengths.  
In the special case of $d=4$, the solutions will be completely
equivalent\footnote{Essentially, the field strength $F_2$ in the $d=4$ 
specialisation of the theory (\ref{sl3rlag}) is acting like the
dualisation double of the field strength $F_1$.} to the dyonic solutions of the theory described by 
(\ref{kkdyonlag}).

For the extremal but non-BPS black holes in the theory
described by (\ref{sl3rlag}), 
we find that if the charges $(Q_1,Q_2)$ and $(\wtd Q_1,\wtd Q_2)$ 
of two such black holes are unequal\footnote{It will always be
understood that we are restricting attention to cases where the {\it signs}
of the charges will be the same for the two black holes.  Obviously,
there is little of interest to discuss if the charges of the two black
holes are opposite, since then there would be an attractive electrostatic 
force that would reinforce the attractive gravitational force, and there
could never be any question of achieving a zero-force balance.  We 
shall always assume, without losing generality for the cases that
are interesting to discuss, that the charges are all positive.},
then the force between the two is always repulsive.\footnote{To be more
precise, the force between them is repulsive provided that
$(\wtd Q_1,\wtd Q_2)$ is not a multiple of $(Q_1,Q_2)$, \ie that 
$(\wtd Q_1, \wtd Q_2) \ne (k\, Q_1, k\, Q_2)$ for any constant $k$.}  
(See \cite{luwazh} for some related discussion of black holes
that repel one another.)
Combining this with the fact that the force between identical black holes is attractive if they are non-extremal 
(\ie sub-extremal), we conclude, by continuity, that that there should exist non-extremal black holes with unequal 
charges for which the parameters can be tuned so that a zero-force 
condition holds. 
 It does not appear to be possible to obtain explicit
analytic formulae for the general parameter choices that achieve such a 
zero-force condition.  However, we are able to solve numerically to
find examples of non-identical non-extremal black holes 
which have no long-range force between them.  We also find an explicit
analytical formula for the zero force condition for a restricted 
sub-family of these black holes.
This balancing of forces is interesting and may hint at a deeper structure, something which we would like to better understand.
We shall come back to this point in the Conclusions.

  The equations for black hole solutions in the theory 
described by the Lagrangian (\ref{sl3rlag}) can be recast as $SL(3,R)$
Toda equations.  In \cite{luyang}, extensions of the
theory that give rise to the $SL(n,R)$ Toda equations were also considered,
and the black hole solutions were constructed.  For the $SL(n,R)$  case there are now $(n-1)$
field strengths, each carrying an electric charge, and $(n-2)$
dilatonic scalar fields.  In the present work we 
study the black holes in the $SL(4,R)$ Toda theory, and show that analogous results
to those we found for the $SL(3,R)$ Toda black holes arise here also.

The paper is organised as follows.  In section 2 we discuss the general 
non-extremal 2-charge black holes of the $SL(3,R)$ Toda theory, and
obtain the expression for the force between two well-separated such
black holes.  Specialising to the case where the two black holes are extremal,
we show that the force between them is always non-negative; that is, it
is always either repulsive or else it is zero.  Specifically, the force
vanishes if and only if the charges $(Q_1,Q_2)$ and $(\wtd Q_1,\wtd Q_2)$ of 
the black holes are proportional, \ie if $(\wtd Q_1,\wtd Q_2)=
(k\, Q_1, k\,Q_2)$.  The case $k=1$ corresponds to
identical black holes.  We then turn to the consideration of two
non-extremal black holes.  If they are identical, we find that the 
force is negative (\ie attractive), in line with standard results in the
literature.  We then show, by means of numerical studies, that for two 
non-identical non-extremal black holes, it is possible to choose the
mass and charge parameters so that there is zero force between them.  We
also obtain an explicit analytical formula characterising the
zero-force condition for a special subset of the non-extremal black holes.

In section 3, we extend our discussion to black hole solutions of the
$SL(4,R)$ Toda theory.  
We demonstrate that the same general
features we found for the $SL(3,R)$ Toda black holes occur 
in this case also.  Namely, the force between two non-identical
extremal black holes is in general repulsive, becoming zero when the
black holes are identical and in certain other special cases.  Furthermore, 
we show with numerical examples that one can tune the parameters of
two non-extremal black holes such that the force between them vanishes.

In section 4 we present our conclusions and further discussion.  Appendix
A contains some details of the calculation of the ADM mass and the 
scalar and electric charges, and gives the relation between our 
normalisations for these quantities and the normalisations in 
\cite{Heidenreich:2020upe}.  In appendix B, we present a calculation of the
force between widely separated BPS black holes in a wide class of
theories, to illustrate some salient features of the differences
between the BPS and the non-BPS extremal black holes.

\section{$SL(3,R)$ Toda Black Holes}

 The $SL(3,R)$ Toda Lagrangian (\ref{sl3rlag}) admits 2-charge static, 
asymptotically-flat black-hole solutions, given by \cite{luyang}
\bea
ds^2 &=& -(H_1\, H_2)^{-\fft12} \, f \, dt^2 + 
 (H_1\,H_2)^{\fft{1}{2(d-3)}}\, \Big( f^{-1}\, dr^2 + r^2\, 
  d\Omega_{d-2}^2\Big)\,,\nn\\
\phi&=& \ft12 \sqrt{\fft{3(d-2)}{2(d-3)}}\, \log\Big(\fft{H_1}{H_2}\Big)\,,
\qquad f= 1 -\fft{m}{r^{d-3}}\,,\nn\\
A_1&=& \sqrt{\fft{d-2}{d-3}}\, 
 \fft{1-\beta_1\, f}{\sqrt{\beta_1\,\gamma_2}\,H_1}\, dt\,,\qquad
A_2\sqrt{\fft{d-2}{d-3}}\,
 \fft{1-\beta_2\, f}{\sqrt{\beta_2\,\gamma_1}\,H_2}\, dt\,,\nn\\
H_1 &=& \gamma_1^{-1}\, (1-2\beta_1\, f +\beta_1\,\beta_2\, f^2)\,,\qquad
H_2=\gamma_2^{-1}\, (1-2\beta_2\, f +\beta_1\,\beta_2\, f^2)\,,\nn\\
\gamma_1 &=& 1-2\beta_1 + \beta_1\, \beta_2\,,\qquad
\gamma_2=1-2\beta_2 + \beta_1\, \beta_2\,,\label{sl3rsol}
\eea
where $m$, $\beta_1$ and $\beta_2$ are constants that parameterise the
mass and the two electric charges.
Adopting convenient normalisation, which we define in appendix A, 
the mass $M$, the scalar charge $\Sigma$, and the two 
physical 
electric charges 
$Q_1$ and $Q_2$ of the black hole solutions are given by 
\bea
M&=& \fft{2(1-\beta_1)(1-\beta_2)(1-\beta_1\, \beta_2)\, m}{
        \gamma_1\,\gamma_2} =2\Big(\fft1{\gamma_1} +\fft1{\gamma_2}\Big)\,
   (1-\beta_1\,\beta_2)\, m\,,\nn\\
\Sigma &=& \fft{\sqrt3\, (\beta_1-\beta_2)\,(1-\beta_1\,\beta_2)\, m}{
                \gamma_1\,\gamma_2}\,,\nn\\
Q_1 &=& \fft{\sqrt{2\beta_1\, \gamma_2}\, m}{\gamma_1}\,,\qquad
 Q_2= \fft{\sqrt{2\beta_2\, \gamma_1}\, m}{\gamma_2}\,. \label{rescaled}
\eea
It will be understood in all that follows that the charges $Q_1$ and
$Q_2$ are assumed to be non-negative, 
so that the gauge force between two black holes will always be repulsive \footnote{ \label{asympdil} In this section,  we always set the 
asymptotic (i.e.  $r \rightarrow \infty$) value of the dilaton to zero in 
the mass, electric charges and scalar charge.  The reason for this choice is because the Lagrangian in \eqref{sl3rlag} has a shift symmetry under $\phi \rightarrow \phi + {\bar\phi}$, with $F_1 \rightarrow e^{-a{\bar\phi}/2}F_1$  and $F_2 \rightarrow e^{a{\bar\phi}/2}F_2$.  Because of this symmetry, we can scale the physical charges as $Q_1\rightarrow e^{a{\bar\phi}/2}Q_1$ and $Q_2\rightarrow e^{-a{\bar\phi}/2}Q_2$,  and using this scaling,  any expression with zero asymptotic value for the dilaton can be dressed with 
a non-zero asymptotic value.  Moreover, since this is a symmetry,  
it will not change any conclusion we draw for the long-range force.}.

  A different parameterisation, which will prove useful for some 
purposes, is provided by expressing $\beta_1$ and $\beta_2$ in terms of 
two new parameters
$p$ and $q$, with
\bea
\beta_1= \fft{(p-m) q}{(q+m) p}\,,\qquad
      \beta_2=  \fft{(q-m) p}{(p+m) q}\,.\label{betapq}
\eea
In terms of $p$ and $q$, the mass, scalar charge and electric charges
in (\ref{rescaled}) become
\bea
M = p + q, \quad \Sigma = \fft{\sqrt{3}}{2} (p - q), 
\quad Q_1 = \left[\fft{p(p^2-m^2)}{(p+q)}\right]^{\fft12}, 
\quad Q_2 = \left[\fft{q(q^2-m^2)}{(p+q)}\right]^{\fft12}\,.
\label{sl3charges}
\eea
It is useful also to note that the Hawking temperature of the black hole
is given by \cite{luyang}
\bea
T=\fft{(d-3)}{4\pi r_+}\, (\gamma_1\,\gamma_2)^{\fft{d-2}{4(d-3)}}\,,
\label{sl3rT1}
\eea
where $r_+$ is the radius of the outer horizon, given by $r_+^{d-3}=m$.
In terms of the parameters $p$ and $q$ introduced in eqn (\ref{betapq}),
this becomes
\bea
T= \fft{(d-3)\, m}{4\pi}\, \Big[\fft{4(p+q)^2}{p q (p+m)^2 (q+m)^2}
  \Big]^{\fft{d-2}{4(d-3)}}\,.\label{sl3rT2}
\eea

As we remarked earlier, in four dimensions the black holes with two
electric charges we are discussing here are equivalent to the KK dyonic
black hole solutions of the theory given in eqn (\ref{kkdyonlag}), with
the electric and magnetic charges of the single field strength $F$ in
(\ref{kkdyonlag}) corresponding to the two electric charges $Q_1$ and
$Q_2$ in eqn (\ref{sl3charges}). In fact the parameterisation using 
$p$ and $q$ as in eqn (\ref{sl3charges}) is precisely\footnote{Note, however, that our extremality parameter $m$ is twice that used in \cite{Cremonini:2021upd}.} the one used
in \cite{Cremonini:2021upd}, where the force between identical Kaluza-Klein
dyons was discussed.

   The force between two well-separated such black holes is given, up to an
overall scale that we suppress for now, by $F=\cF\, r^{-(d-2)}$, where
\bea
\cF= Q_1\, \wtd Q_1 + Q_2\,\wtd Q_2 - \Sigma\, \wtd\Sigma -
   \ft14 M\, \wtd M\,,\label{cF}
\eea
where the untilded and tilded quantities refer to the two black holes,
with the untilded quantities being given in terms of parameters 
$(m,\beta_1,\beta_2)$, and the tilded quantities in terms of 
parameters $(\tilde m, \tilde \beta_1,\tilde \beta_2)$.  For the explicit 
relation of our masses and charges to those defined 
in \cite{Heidenreich:2020upe} we refer the reader to appendix A.

We now examine the nature of the force between the black holes in two
cases; firstly, for extremal black holes, and then for non-extremal
black holes.

\subsection{Extremal $SL(3,R)$ Black Holes}

One way of taking the extremal limit of the $SL(3,R)$ black hole solutions 
is by setting \cite{luyang}
\bea
\beta_1 &=& 1- q_1^{-\fft23}\, q_2^{-\fft23}\, 
             (q_1^{\fft23} + q_2^{\fft23})^{\fft12}\, m +
              q_1^{-\fft23}\, q_2^{-\fft43}\, m^2\,,\nn\\
\beta_2 &=& 1- q_1^{-\fft23}\, q_2^{-\fft23}\,
             (q_1^{\fft23} + q_2^{\fft23})^{\fft12}\, m +
              q_1^{-\fft43}\, q_2^{-\fft23}\, m^2\,,\label{betaext}
\eea
in terms of two new charge parameters $q_1$ and $q_2$.  The
extremal limit is then attained by sending $m\longrightarrow 0$.  
In this limit, the mass, scalar and 
electric charges defined in (\ref{rescaled}) become
\bea
M_{\rm ext}&=& (q_1^{\fft23} + q_2^{\fft23})^{\fft32}\,,\nn\\
\Sigma_{\rm ext} &=& \fft{\sqrt3}{2}\, (q_1^{\fft23}-q_2^{\fft23})\, 
        (q_1^{\fft23} + q_2^{\fft23})^{\fft12}\,,\nn\\
Q_1^{\rm ext} &=& q_1\,,\qquad Q_2^{\rm ext} = q_2\,.\label{extremal}
\eea
In this extremal limit, the (double) horizon is at $r=0$. Inserting
eqns (\ref{betaext}) into the expression (\ref{sl3rT1}) for the
Hawking temperature gives
\bea 
T=\fft{(d-3) m}{4\pi}\, \Big[\fft2{q_1 q_2}\Big]^{\fft{d-2}{2(d-3)}}
+ {\cal O}(m^2)\,,
\eea
which goes to zero, as one would expect, in the extremal limit
$m\longrightarrow 0$.

It is straightforward to verify that if two such extremal 
black holes have charges that are multiples of one another, \ie~if
\bea
(\tilde q_1,\tilde  q_2) = (k\, q_1, k\, q_2)\,,\label{eqq}
\eea
then the force between them, given by substituting eqns 
(\ref{extremal}) into eqn (\ref{cF}), is zero.  This includes the special
case $k=1$, corresponding to two identical extremal black holes.

Next, we would like to examine the general case of two non-identical extremal black holes.
The solutions may be reparameterised in terms of $(q,\theta)$ and 
$(\tilde q, \tilde\theta)$, where 
\bea
q_1= q^3\, \cos^3\theta\,,\qquad q_2=q^3\,\sin^3\theta\,,\qquad
\tilde q_1= \tilde q^3\, \cos^3\tilde\theta\,,\qquad 
\tilde q_2=\tilde q^3\,\sin^3\tilde\theta\,.\label{qtheta}
\eea
With the understanding that the charges are all non-negative, we see that
we must have
\bea
0\le \theta\le \ft12\pi\,,\qquad 0\le\tilde\theta\le \ft12\pi\,.
\label{thetalim}
\eea
Plugging (\ref{qtheta}) into (\ref{extremal}), and then these into
the expression (\ref{cF}) for the force coefficient, we find
\bea
\cF = q^3\, \tilde q^3\, G\,,
\eea
where
\bea
G= \cos^3\theta\,\cos^3\tilde\theta + \sin^3\theta\, \sin^3\tilde\theta 
-\ft34 \cos2\theta\,\cos2\tilde\theta -\ft14\,.\label{Gdef}
\eea
As we shall now show, $G$ is non-negative when $\theta$ and
$\tilde\theta$ lie anywhere in the square defined by (\ref{thetalim}).

  First, we define $t=\tan\ft12\theta$ and $\tilde t=\tan\ft12\tilde\theta$,
so we have
\bea
\sin\theta= \fft{2t}{1+t^2}\,,\qquad \cos\theta= \fft{1-t^2}{1+t^2}\,,
\qquad
\sin\tilde\theta= \fft{2\tilde t}{1+\tilde t^2}\,,\qquad 
\cos\tilde\theta= \fft{1-\tilde t^2}{1+\tilde t^2}\,.
\eea
In view eqn (\ref{thetalim}) we have $0\le t\le1$ and $0\le\tilde t\le 1$.  
Thus we can parameterise $t$ and $\tilde t$ as
\bea
t=\fft{a}{1+a}\,,\qquad \tilde t = \fft{b}{1+b}\,,
\eea
with
\bea
0\le a\le\infty\,,\qquad 0\le b\le \infty\,.\label{abrange}
\eea
Plugging these substitutions into the definition of $G$ in (\ref{Gdef}), 
we find
\bea
G= \fft{2(a-b)^2\, P(a,b)}{(1+2a + 2a^2)^3\, (1+2b + 2b^2)^3}\,,\label{GP}
\eea
where 
\bea
P(a,b)&=& 3(a+b)^2 + 6 (a+b)(a^2 + 4 a b + b^2) +
  2( a^4 + 20 a^3 b + 36 a^2 b^2 + 20 a b^3 + b^4)\nn\\
&& +
12 a b (a+b) (a^2 + 4 ab + b^2) + 12 a^2 b^2 (a+b)^2\,.
\eea
Since all the coefficients in $P(a,b)$ are positive, it follows that
$P(a,b)\ge0$ for all $a$ and $b$ in the range (\ref{abrange}), and thus we
see from eqn (\ref{GP}) that $G\ge0$.  

Thus, we have shown that the
force between any two extremal $SL(3,R)$ Toda black holes is always
non-negative, and that it is stricly positive (\ie repulsive) provided
that the charges of the two black holes are not proportional
(\ie provided that eqn (\ref{eqq}) is not satisfied for any constant $k$).

  It is worth remarking that taking the extremal limit is somewhat more
straightforward
in the parameterisation using $p$ and $q$ as in eqn (\ref{betapq}),
since one can now simply set $m=0$ in the expressions (\ref{sl3charges})
for the mass, scalar charge and electric charges, without the need
for taking a delicate limit. Indeed, 
at extremality
the electric charges
$q_1$ and $q_2$ in eqns (\ref{betaext}) are given in terms of $p$ and $q$
by
\bea
q_1 = \fft{p^{\fft32}}{(p+q)^{\fft12}}\,,\qquad
q_2 = \fft{q^{\fft32}}{(p+q)^{\fft12}}\,,
\eea
as can be seen directly from eqns (\ref{sl3charges}).  Note that in
this parameterisation one can directly see that the Hawking temperature
of the black hole beccomes zero in the extremal limit, by setting $m=0$
in eqn (\ref{sl3rT2}).

\subsection{Non-Extremal $SL(3,R)$ Black Holes}

  Now consider non-extremal black holes, characterised by parameters
$(m,\beta_1,\beta_2)$ and $(\tilde m, \tilde\beta_1,\tilde\beta_2)$
respectively. 
The general expression for the force between two distinct non-extremal black holes is the following,
\bea
{\cal F} &=& \fft{m{\td m}}{\gamma_1\gamma_2{\td\gamma_1}{\td\gamma_2}}
\Big[\, 2\,(\beta_1{\td\beta_1})^\fft12(\gamma_2{\td\gamma_2})^\fft32 
+ 2\,(\beta_2{\td\beta_2})^\fft12(\gamma_1{\td\gamma_1})^\fft32 \nn\\
&& \hspace{1.8cm}-3\,(\beta_1-\beta_2)
  (1-\beta_1\beta_2)({\td\beta}_1-{\td\beta}_2)
    (1-{\td\beta}_1{\td\beta}_2) \nn\\
&& \hspace{1.8cm}  -\, (1-\beta_1)(1-\beta_2)
(1-\beta_1\beta_2)(1-{\td\beta}_1)(1-{\td\beta}_2)
  (1-{\td\beta}_1{\td\beta}_2) \Big]\,.\label{neforce}
\eea
It can be verified easily that if we consider
the case where
\bea
(\tilde\beta_1,\tilde\beta_2) =(\beta_1,\beta_2)\,,\label{betaeq}
\eea
which corresponds to the tilded electric charges being an overall multiple
of the untilded electric charges (see the expressions in eqns (\ref{rescaled})
for $Q_1$ and $Q_2$), then the force between the non-extremal
black holes will be attractive, with $\cF$ given simply by
\bea
\cF = -m\, \tilde m\,.\label{neeq}
\eea
The special case of identical non-extremal 
black holes arises when the further condition $\tilde m=m$ is 
imposed.\footnote{It can also be seen that if the black holes obeying
eqn (\ref{betaeq}) are taken to be extremal, by sending $m$ and $\tilde m$
to zero, then the force (\ref{neeq}) 
between them becomes zero, as already noted for
extremal black holes whose charges are proportional. Note that one must
be careful when taking the extremal limit in this parameterisation, since
the $\beta_i$ parameters must be taken to 1 at the same time, 
as seen in the limiting procedure in eqn (\ref{betaext}).  In particular,
having imposed the requirement (\ref{betaeq}) one could not take the
$m=0$ extremal limit for the untilded black hole without also taking the
$\tilde m=0$ limit for the tilded black hole, since otherwise the tilded
charges $\wtd Q_i$ would become infinite.} 

Recall that for the case of two \emph{distinct} extremal black holes, we have shown that the long range force is always positive.
Thus, since the force must presumably be a continuous
function of the parameters, we conclude that for a general pair of non-extremal
black holes, there must exist some
choices of the $m$ and $\beta_i$ parameters which yield a zero-force 
condition.  
Indeed, here is a 
numerical example:

   We write $\beta_1$ and $\beta_2$ as in eqn (\ref{betaext}), but we do 
{\it not} send $m$ to zero.\footnote{At this stage,
therefore, we just have a reparameterisation of non-extremal black 
holes in terms of $m$, $q_1$ and $q_2$ rather than $m$, $\beta_1$ and
$\beta_2$.  It is a convenient repararmeterisation to adopt here since
it allows us explore the situation where the black hole is
becoming close to extremality, by taking $m$ to be fairly small (in 
comparison to $q_1$ and/or $q_2$).  Note that $q_1$ and $q_2$ are 
{\it not} simply multiples of the physical charges $Q_1$ and $Q_2$,
except in the actual extremal limit where $m\longrightarrow 0$.} 
We do likewise for $\tilde\beta_1$ and
$\tilde\beta_2$ (taking, for convenience, $\tilde m=m$).  Making a
specific choice for the untilded and tilded $q_i$ parameters, namely
\bea
q_1=1\,,\qquad q_2=2\,,\qquad \tilde q_1=3\,,\qquad \tilde q_2=7\,,
\eea
we then look at the numerical value of the force coefficient $\cF$, as 
a function of $m$. We find that, approximately,
\bea
m> 0.066\qquad \hbox{implies} \qquad \cF <0\,,\nn\\
m< 0.066 \qquad\hbox{implies} \qquad \cF >0\,.
\eea
In other words, for this example of two black holes whose charges are
not simply an overall multiple of one-another, the force between them is
repulsive when they are sufficiently close to being extremal, but it
becomes attractive when they are taken to be sufficiently far from
extremality.  The zero-force condition arises when the non-extremality
parameter $m$ is roughly equal to $0.066000632$.

As another example, if we take
\bea
q_1=1\,,\qquad q_2=2\,,\qquad \tilde q_1=1\,,\qquad \tilde q_2=3\,,
\eea
then the crossover between repulsion and attraction occurs when the
non-extremality parameter $m$ is approximately
\bea
m= 0.2073984664\,.
\eea
Note that it is necessary to check that the constants $\beta_i$, 
$\tilde\beta_i$, $\gamma_i$ and $\tilde\gamma_i$ are all non-negative, in order
to ensure that the black holes are regular from the horizon to 
asymptotic infinity.  In the examples above, these conditions are indeed
satisfied.

  We may also give an explicit construction of a special family of 
non-identical non-extremal black holes that obey the zero-force 
condition.  Using the parameterisation in terms of $p$ and $q$ as
in eqns (\ref{betapq}), the expression (\ref{neforce}) for the force
between two non-extremal black holes becomes
\bea
{\cal F} &=& -\fft14 \bigg[(p+q)({\td p}+\td{q}) + 
   3 (p-q)(\td{p}-\td{q})\bigg]\nn\\
&& + \left[\fft{p{\td p} (p^2 - m^2)({\td p}^2 -
{\td m}^2)}{(p+q)({\td p}+{\td q})}\right]^{\fft12} +
\left[\fft{q{\td q} (q^2 - m^2)({\td q}^2 -
  {\td m}^2)}{(p+q)({\td p}+{\td q})}\right]^{\fft12}\,.\label{KKF}
\eea
As we saw previously, for two identical non-extremal black holes the
force becomes ${\cal F} = -m^2$. 
Note that in order to avoid imaginary charges and negative mass, 
we should restrict $p$ and $q$ such that $p\ge m$ and $q\ge m$. 

   Consider now the following specialisation.  Take
\bea
\tilde m=m\,,\qquad \tilde p=q\,,\qquad \tilde q =p\,,\label{pqswitch}
\eea
for which the expression (\ref{KKF}) becomes
\bea
\cF = \fft{2\sqrt{pq}\, \sqrt{(p^2-m^2)(q^2-m^2)}}{p+q} +
    \ft12 (p^2+q^2-4pq)\,.\label{KKFspec}
\eea
For $p=q$ we find $\cF=-m^2$ (as is to be expected for two identical
non-extremal black holes); the second term is negative, and outweighs
the first term. If instead $p\ne q$, then 
under certain circumstances the second term in (\ref{KKFspec}) will
be positive, and so $\cF$ will be positive.  Thus within this
considerably simplified class of solutions, we can find 
explicit expressions for intermediate 
cases that achieve a zero-force condition:

   Writing
\bea
p= x\, q\,,
\eea
we can solve for the ratio $q^2/m^2$ that makes $\cF$ in eqn
(\ref{KKFspec}) vanish.  This gives
\bea
\fft{q^2}{m^2} = -\fft{4[2x(1+x^2)\pm \sqrt{x}\, (1+x)
   \sqrt{1-4x+10x^2-4x^3+x^4}]}{(x-1)^2(1-4x-6x^2-4x^3+x^4)}\,.\label{qsq0}
\eea
We must choose $x$ so that we have $q^2-m^2>0$, and also
$p^2-m^2>0$ (\ie $q^2\, x^2-m^2>0$).  This implies we should make the
upper sign choice in eqn (\ref{qsq0}), and so, defining
\bea
W(x) \equiv -\fft{4[2x(1+x^2) + \sqrt{x}\, (1+x)
   \sqrt{1-4x+10x^2-4x^3+x^4}]}{(x-1)^2(1-4x-6x^2-4x^3+x^4)}\,,\label{Wdef}
\eea
we shall have
\bea
\fft{q^2}{m^2}= W(x)\,,\qquad 
\fft{p^2}{m^2}= x^2\, W(x) = W\big(\fft1{x}\big)\,.\label{pqres}
\eea
Since sending $x\longrightarrow x^{-1}$ exchanges the roles of $p$ and $q$,
which corresponds to the symmetry of the $SL(3,R)$ theory under the
reflection of the the $SL(3,R)$ Dynkin diagram,
we can, without loss of generality, restrict attention to taking $x$, which 
must be positive, to lie in the interval $1\le x\le\infty$.

In the interval $1\le x\le\infty$ it is evident from eqns (\ref{pqres})
that the conditions $p^2\ge m^2$ and $q^2\ge m^2$ will be satisfied if
$W(x)\ge1$, and one can see from eqn (\ref{Wdef}) that this will hold
if 
\bea
1<x<x_+\,,\qquad  \hbox{where}\quad x_+\approx 5.27451\,,\label{xint}
\eea
with $x_+$ being the larger of the two real roots of $1-4x-6x^2-4x^3+x^4=0$.
The function $W(x)$ is strictly greater than 1 for all $x$ in the interval
(\ref{xint}), except when $x=2+\sqrt3 \approx 3.73205$, for which
$W$ becomes equal to 1.

As a concrete example, if we take $x=3$ then
\bea
q= m\, \sqrt{\fft{15+2\sqrt{39}}{23}}\,,\qquad
p=3 m\, \sqrt{\fft{15+2\sqrt{39}}{23}}\,.
\eea
This satisfies all the necessary constraints, and indeed gives $\cF=0$.
For this particular example, the temperatures of the two 
non-extremal black hole solutions are the same (since, as one can see from
eqn (\ref{sl3rT2}), if we denote the temperature by $T(p,q,m)$ then
we have $T(p,q,m)=T(q,p,m)$).  By contrast, in each of 
the previous numerical
examples we presented, the temperatures were unequal for the 
two non-extremal black holes for which a no-force condtion held.
While the equality of the temperatures in the example in eqn (\ref{pqswitch})
is due to the symmetrical parameter choices of these particular 
solutions, it may also 
hint at the 
existence of a new multi-charge black hole. 
We will return to this point in the Conclusions.
Finally, we anticipate that there shouldn't be any obstruction to 
finding a vanishing force for parameter choices corresponding 
to properly quantized physical charges, once one is more careful 
with normalizations and properly reinstates units.

\section{$SL(4,R)$ Toda Black Holes}

 $SL(n,R)$ Toda black holes are discussed in \cite{luyang}, and additional
explicit details are given for the $SL(4,R)$ case.  The Lagrangian for
the $SL(n,R)$ case is
given by
\bea
{\cal L} &=& R - \ft12 (\del\vec\phi)^2 -\ft14 \sum_{i=1}^{n-1}
  e^{\vec a_i\cdot\vec\phi}\, F_i^2\,,
\eea
where the $(n-1)$ dilaton vectors $\vec a_i$ satisfy
\bea
\vec a_i\cdot\vec a_i &=& \ft13 (n-2)(n^2+2n+3)\, \fft{d-3}{d-2}\,,\qquad
(\hbox{no sum on}\ i)\,,\nn\\
\vec a_i\cdot\vec a_{i+1} &=& -\ft16 (n^3-n+12)\, \fft{d-3}{d-2}\,,\nn\\
\vec a_i\cdot\vec a_j &=& -\fft{2(d-3)}{d-2}\,,\qquad
i\ne j-1,\ j, \ j+1\label{adots}  \; .
\eea
There are $n-2$ dilatonic scalars, so the dilaton vectors are 
$(n-2)$-component vectors.
For the case of $SL(4,R)$, we can satisfy the conditions (\ref{adots}) 
by choosing
\bea
\vec a_1 =(\sqrt{8\nu},\sqrt{10\nu})\,,\quad
\vec a_2= (-\sqrt{18\nu},0)\,,\quad 
\vec a_3= (\sqrt{8\nu},-\sqrt{10\nu})\,,\quad
\hbox{where}\quad \nu\equiv \fft{d-3}{d-2}\,.
\eea

  The $SL(4,R)$ Toda black hole solutions,  involving 3 field strengths, each carrying an electric charge, and 2 dilatonic scalar fields, were constructed in \cite{luyang}. 
They are given by 
\bea
ds^2 &=& -(H_1 H_2 H_3)^{-\fft15}\, f\, dt^2 +
(H_1 H_2 H_3)^{\fft1{5(d-3)}}\,(f^{-1}\, dr^2 + 
                  r^2\, d\Omega_{d-2}^2)\,,\nn\\
\vec\phi &=& \fft1{10\nu}\, \sum_i \vec a_i\, \log H_i\,,\nn\\
A_1 &=& \sqrt{\fft{3}{5\nu}}\, 
 \fft{1-2\beta_1\, f + \beta_1\, \beta_2\, f^2}{\sqrt{\beta_1\, \gamma_2}\, 
     H_1}\, dt\,,\qquad
A_3 = \sqrt{\fft{3}{5\nu}}\,
 \fft{1-2\beta_1\, f + \beta_3\, \beta_2\, f^2}{\sqrt{\beta_3\, \gamma_2}\,
     H_3}\, dt\,,\nn\\
A_2 &=& \sqrt{\fft{4}{5\nu}}\,
   \fft{1-3\beta_2\, f + \ft32 \beta_2\,(\beta_1+\beta_3)\, f^2 -
\beta_1\,\beta_2\,\beta_3\, f^3}{\sqrt{\beta_2\, \gamma_1\,\gamma_3}\,
   H_2}\, dt\,,
\eea
where
\bea
H_1&=& \gamma_1^{-1}\, (1-3\beta_1\, f + 3\beta_1\,\beta_2\, f^2 - 
    \beta_1\,\beta_2\,\beta_3\,f^3)\,,\nn\\
H_2 &=& \gamma_2^{-1}\, (1-4\beta_2\, f + 3\beta_2\,(\beta_1+\beta_3)\, f^2
 - 4\beta_1\,\beta_2\,\beta_3\, f^3 + \beta_1\, \beta_2^2\, \beta_3\, f^4)
\,,\nn\\
H_3&=& \gamma_3^{-1}\, (1-3\beta_3\, f + 3\beta_2\,\beta_3\, f^2 - 
    \beta_1\,\beta_2\,\beta_3\,f^3)\,,\nn\\
f&=& 1-\fft{m}{r^{d-3}}\,,
\eea
and
\bea
\gamma_1&=& 1-3\beta_1 + 3\beta_1\,\beta_2 -\beta_1\,\beta_2\, \beta_3\,,\nn\\
\gamma_2 &=& 1 -4\beta_2 + 3\beta_2\,(\beta_1+\beta_3)
 - 4\beta_1\,\beta_2\,\beta_3 + \beta_1\, \beta_2^2\, \beta_3\,,\nn\\
\gamma_3&=& 1-3\beta_3 + 3\beta_2\,\beta_3 -\beta_1\,\beta_2\, \beta_3\,.
\eea

 The ADM mass of the $SL(4,R)$ Toda black hole is given in \cite{luyang},
and can be written (in our choice of overall normalisation) as \footnote{Again we choose to set the asymptotic value of the dilaton to zero for reasons mentioned in footnote \ref{asympdil}.}
\bea
M =\Big(\fft{3k_1}{\gamma_1} + \fft{4k_2}{\gamma_2} +
             \fft{3k_3}{\gamma_3} \Big)\, \fft{m}{5}\,,
\eea
where
\bea
k_1&=& 1-\beta_1 -\beta_1\,\beta_2 + \beta_1\,\beta_2\,\beta_3\,,\nn\\
k_2&=& 1-2\beta_2 + 2\beta_1\,\beta_2\,\beta_3 -
   \beta_1\,\beta_2^2\, \beta_3\,,\nn\\
k_3&=& 1-\beta_3 -\beta_2\,\beta_3 + \beta_1\,\beta_2\,\beta_3\,.
\eea
With the overall normalisation we are adopting, the three electric
charges are given by \cite{luyang}
\bea
Q_1 &=& \fft{\sqrt{6\beta_1\,\gamma_2}\, m}{\sqrt5\, \gamma_1}\,,\qquad
Q_2 = \fft{\sqrt{8\beta_2\,\gamma_1\,\gamma_3}\, m}{\sqrt5\,\gamma_2}\,,\qquad
Q_3= \fft{\sqrt{6\beta_3\,\gamma_2}\, m}{\sqrt5\,\gamma_3}\,. 
\eea
The two scalar charges, read off from the coefficients of the
$r^{-d+3}$ terms in the asymptotic expressions for the dilaton fields,
are given, in our normalisation, by the 2-vector
\bea
\vec\Sigma = \fft1{5\sqrt{2\nu}}\, \sum_i \vec a_i \,\ell_i\,,
\eea
and hence
\bea
\Sigma_1 = \fft{2\ell_1 -3\ell_2+2\ell_3}{5}\,,\qquad
\Sigma_2=\fft{\ell_1-\ell_3}{\sqrt5}\,,
\eea
where $\ell_i$ are the coefficients of the $r^{-d+3}$ terms in the 
expressions for $H_i = 1 +\fft{\ell_i}{r^{d-3}}+\cdots$, and are given by
\bea
\ell_1 &=& \fft{3\beta_1\, m}{\gamma_1}\, (1-2\beta_2 + \beta_2\,\beta_3)
\,,\qquad 
\ell_3 = \fft{3\beta_3\, m}{\gamma_3}\, (1-2\beta_2 + \beta_1\, \beta_2)\,,
\nn\\
\ell_2 &=& \fft{4\beta_2\, m}{\gamma_2}\, \big(1-\ft32 (\beta_1+\beta_3) 
  +3\beta_1\,\beta_3 - \beta_1\,\beta_2\,\beta_3\big)\,.
\eea

  The force between two black holes, characterised by untilded and 
tilded parameters $(m,\beta_1,\beta_2,\beta_3)$ and 
$(\tilde m,\tilde\beta_1,\tilde\beta_2,\tilde\beta_3)$, is given by
\bea
\cF= \sum_{i=1}^3 Q_i\, \wtd Q_i - \vec\Sigma\cdot\vec {\wtd\Sigma} 
-\fft14\, M\,\wtd M\,.
\eea
If the two black holes are identical, this gives
\bea
\cF= - m^2\,.
\eea
As expected, this vanishes in the extremal case ($m=0$) and is
negative -- implying an attractive force -- in the sub-extremal case.

   As we also saw in the case of the $SL(3,R)$ Toda system,
studying the force between two non-identical black holes is a lot
more involved. 
However, the logic here is similar to that of the previous section.
As we shall discuss in more detail below, in the case of unequal
extremal black holes the force can be positive,
and although we don't have a general proof in this case, we
can expect that 
it will always be positive, just as we saw for the $SL(3,R)$ examples. 

By continuity, given that the force between identical non-extremal black
holes is negative, we can again expect, just as for $SL(3,R)$ black holes,
that there should exist non-identical pairs of non-extremal $SL(4,R)$ 
black holes for which a zero-force condition holds.  Here, we present
one explicit numerical example.  We consider two non-extremal 
$SL(4,R)$ black holes with parameters
\bea m=\tilde m\,,\qquad
(\beta_1,\beta_2,\beta_3)=(\ft15, \ft13, \ft14)\,,\qquad
(\tilde\beta_1,\tilde\beta_2,\tilde\beta_3)=(\ft15+w, \ft13+w, \ft14+w)\,.
\eea
When $w=0$ the force is just given by $\cF= -m^2$, as mentioned 
previously. We find that the force becomes zero if the parameter $w$ 
is tuned to 
\bea
w \approx 0.03742503\,.
\eea

   In the case of extremal $SL(4,R)$ Toda 
black holes, we can follow
the procedure described in \cite{luyang} for taking the extremal
limit, by writing\footnote{A sign error in \cite{luyang}
in the expression for $\beta_2$ is corrected here.} 
the $\beta_i$ in terms of new parameters $a,b,c$, as follows:
\bea
\beta_1 = 1 - a\, m + a^2\, b\, m^2\,,\quad
\beta_2= 1 -a\, m + \ft12 a^3\, (c-1)\, m^3\,,\quad
\beta_3= 1 - a\, m - a^2\, b\, m^2\,,\label{betaabc}
\eea
and then sending $m$ to zero:  Before taking this limit, the temperature
is given by \cite{luyang}
\bea
T= \fft{(d-3)}{4\pi r_+}\, (\gamma_1\gamma_2\gamma_3)^{\fft{d-2}{10(d-3)}}\,,
\eea
which implies, using the parameterisation in eqn (\ref{betaabc}), that
\bea
T=\fft{(d-3) m}{4\pi}\, [a^{10}\, (c^2-9b^2)(3+3b^2-2c)]^{\fft{d-2}{10(d-3)}}
+ {\cal O}(m^2)\,,
\eea
which indeed vanishes when $m=0$.
   
Using (\ref{betaabc}) and then sending $m$ to zero 
gives charges and mass as follows:
\bea
Q_1 &=& \fft{\sqrt{6(3+3b^2 -2c)}}{\sqrt5\, a\, (c-3b)}\,,\qquad
Q_2 = \fft{2\sqrt2\, \sqrt{c^2-9b^2}}{\sqrt5\, a\, (3+3b^2-2c)}\,,\qquad
Q_3 = \fft{\sqrt{6(3+3b^2 -2c)}}{\sqrt5\, a\, (c+3b)}\,,\nn\\
\Sigma_1 &=& \fft{6 (9b^2 - 18b^4 + 6c + 9b^2\, c - 7c^2 +c^3)}{
5a\, (3+3b^2 -2c)(c^2-9b^2)}\,,\nn\\
\Sigma_2 &=& \fft{6 b\, (3-c)}{\sqrt5\, a \,  (c^2-9b^2)}\,,
\nn\\
M &=& \fft{4(36 b^2\, c + 9c -54b^2 -27 b^4 - 3c^2 -c^3)}{5 a\,
   (3+3b^2 -2c)(c^2-9b^2)}\,.
\eea
The parameters $b$ and $c$ are constrained by the requirements that
\bea
0\le b\le 1\,,\qquad 3b \le c\le \ft32 (1+b^2)\,,
\eea
where we have, without loss of generality, required $b$ to be non-negative 
(there is a symmetry of the solution, corresponding to reflecting the 
$SL(4,R)$ Dynkin diagram, under sending $b\longrightarrow -b$ and exchanging
the ``1'' and ``3'' labels on the electric charges).  
It will be 
convenient to parameterise the allowed ranges of values for $b$ and
$c$ in terms of two parameters $x$ and $y$ that can independently
range over $0\le x\le \infty$ and $0\le y\le\infty$, by writing
\bea
c=3b+ \fft{3(1-b)^2\, y}{2(1+y)}\,,\qquad b= \fft{x}{1+x}\,.\label{bcsub}
\eea

  From looking at numerous numerical examples, it would seem that the
force between any two unequal-charge extremal $SL(4,R)$ Toda black holes
is positive (\ie replulsive).  Since we were able to prove this
explicitly for the analogous $SL(3,R)$ case, it 
would seem likely that the force will be always repulsive in this case too.
  Although we have not constructed an explicit proof in general, we 
are able to show that when the parameters of the 
two extremal black holes are close to one another, the force is always
repulsive.  Specifically, we may consider the situation where
\bea
\tilde a =a\,,\qquad \tilde b = b + \epsilon_1\,,\qquad
  \tilde c= c + \epsilon_2\,,
\eea
and then look at the expression for the long-range force at leading order
in the small quantities $\epsilon_1$ and $\epsilon_2$   (since the
parameters $a$ and $\tilde a$ appear only as overall scaling factors in the
expressions for masses and charges, there is no loss of generality,
from the point of view of establishing a positivity result, in simply
taking $\tilde a$ to be equal to $a$).  The leading order terms in the
expression for the force arise at the quadratic order in the
$\epsilon$ parameters, so $\cF$ takes the form
\bea
\cF = h_{ij}\, \epsilon_i\, \epsilon_j + {\cal O}(\epsilon^3)\,.
\eea
We find that
\bea
\det(h_{ij}) = \fft{64(1+x)^{10}\, (1+y)^5\, (1+4x + 2x^2 + 2y+ 4 x y
  +2 x^2\, y)}{225 a^4\, y^2\, (4x+4x^2 + y + 4x y + 4 x^2y)^2}\,,
\eea
which is non-negative since $0\le x\le\infty$ and
$0\le y\le\infty$. We also find that $\sum_i h_{ii}$ is manifestly
non-negative.  Thus, it must be that the two eigenvalues of $h_{ij}$
are non-negative, and so at least to quadratic order in 
perturbations around the case of identical extremal black holes,
the force is always repulsive, as anticipated.

\section{Discussion and Conclusions}

Motivated quite naturally by studies of the WGC, there has been recent interest in exploring the 
relation between black hole extremality bounds and zero-force conditions. 
In particular, the RFC encodes the idea that gravity is the weakest force by requring (in its simplest form) 
the existence of self-repulsive states.
However, while the WGC has been examined extensively -- and partial proofs have appeared in a variety of contexts -- the RFC is thus far less studied and much less understood.  
While the two conjectures are identical in two-derivative theories that contain only gravity and electromagnetic forces (where a charged state with $Q>M$ will clearly repel itself), they are known to be distinct in the presence of scalar fields, which can mediate new long-range interactions. 
In this context, the fact that 
long-range forces between identical extremal black holes vanish independently of the complexity of the matter sector (which may include a variety of scalar and gauge fields) is quite non-trivial.

The situation is more complicated in the presence of higher derivative corrections, 
where there are examples of theories in which forces between extremal black holes don't have definite signs, and of directions in charge space with no self-repulsive states \cite{Cremonini:2021upd}.
Indeed, the analysis  of \cite{Cremonini:2021upd} suggests that at least 
the simplest versions of the RFC may be violated by the black hole spectrum -- in stark contrast with the WGC, 
which could in principle be satisfied entirely by black holes. 
While this doesn't rule out the RFC -- it simply requires the existence of self-repulsive fundamental particles -- it does highlight the fact that the conjecture is fundamentally different from the WGC, and raises the question of 
to what extent long-range forces can be used to constrain low-energy EFTs and how much non-trivial information they encode.
Higher derivative corrections to long-range forces were also studied in \cite{Ortin:2021win}, which
computed $\alpha^\prime$ corrections to families of heterotic multi-center black hole solutions. 
Interestingly, in the cases studied in \cite{Ortin:2021win}, the zero-force condition between extremal black holes was preserved even 
in the presence of higher derivatives, for both supersymmetric and non-supersymmetric solutions. 
Once again this raises the question of how to interpret more generally the implications of the balancing of forces.

Motivated by these issues, in this paper we have extended the studies of long-range forces to black holes that are \emph{not} identical. 
We were primarily interested in whether one could identify any generic features in the behavior of the corresponding forces,
and perhaps 
shed light on what properties of black hole solutions can be captured by treating them as widely separated point particles, 
working essentially in a non-relativistic, weak field limit.  
As we have seen, the behavior of the forces is quite rich and leads to a number of novel results.
In particular, any pair of (non-BPS) extremal black holes that are not identical (up to overall scaling) 
repel one another.  We constructed a proof for the case of $SL(3,R)$ Toda black
holes,  and similar considerations seemingly apply to the more complicated
example of $SL(4,R)$ Toda black holes.
Since we have restricted our attention to somewhat simple classes of 
black holes which carry only two (electric) charges 
(in the $SL(3,R)$ Toda case), 
a natural next step is to extend our analysis to the most general 
class of STU black holes in four dimensions, to see whether the features we 
have identified here persist or not for broader classes of solutions, 
including magnetic charges. 
Work in this direction is in progress \cite{crcvposa}.

We have also identified pairs of non-extremal, non-identical black holes which obey a zero-force condition. 
While such balancing of forces is well-known in the context of 
supersymmetric BPS black holes in supergravity, where it holds even for black holes that carry different charges, 
it is unexpected here. Moreover, for multicenter BPS black holes the cancellation of the forces holds for arbitrary relative positions of the centers.
On the other hand, in the theories we have examined the zero-force condition 
applies only to well separated black holes (\ie in the long-range limit) and 
for specific choices of parameters. 
Nonetheless, we wonder whether it may indicate the existence of 
new multi-center black hole solutions (perhaps at nearby points in phase space), 
especially for cases where the two black holes have the same temperature. 
Indeed, it would be valuable to better understand to what extent no-force conditions such as the ones we have found in this paper
can be used as a diagnostic tool for identifying and generating new solutions, aided by suitable constraints on \emph{e.g.}
the temperature, regularity properties and so on. 
It would also be interesting to understanding this logic in light of the results of \cite{Ortin:2021win}, where both supersymmetric and non-supersymmetric black holes, corrected by higher derivatives, satisfy a no-force condition.

A more challenging question is to understand how to connect our results to expectations from the RFC. 
As we have seen, in the theories we have studied distinct extremal black holes repel. 
If this observation is a generic feature of top-down theories that support broader classes of black hole solutions, it may help 
 identify the kinds of repulsive multi-particle states that the strong version of the RFC would call for.
We leave these questions to future work.

\section*{Acknowledgements}

We are grateful to Tomas Ort\'\i{}n and Timm Wrase for useful conversations.
S.C.~is supported in part by NSF grant PHY-1915038. 
M.C.~is supported in part by DOE 
Grant Award de-sc0013528 and the Fay R. and Eugene L. Langberg Endowed Chair.
C.N.P.~is supported in part by DOE grant DE-FG02-13ER42020.

\appendix

\section{Mass and Charges for Static Black Holes}

\subsection{Mass}

  The static $d$-dimensional black-hole metrics can be written in the form
\bea
ds^2= -u\, f\, dt^2 + u^{-\fft1{d-3}}\, \big( f^{-1}\, dr^2 + 
r^2 d\Omega_{d-2}^2\big)\,,\label{statmet}
\eea
where $u$ and $f$ are functions only of $r$.

  We can expand the metric around Minkowski spacetime by introducing 
the coordinates
\bea
x = t\,,\qquad x_i= r\, n_i\,,\qquad \hbox{where}\qquad n_i\, n_i=1\,,
\eea
and the unit vector $n_i$ is parameterised in terms of the $(d-2)$ angular 
coordinates on the unit $(d-2)$-sphere.  We shall have $dn_i\, dn_i=
d\Omega_{d-2}^2$.  We can now write the metric (\ref{statmet}) as follows:
\bea
ds^2 &=& - uf (dx^0)^2+ u^{-\fft1{d-3}}\, (f^{-1} -1)\, dr^2 +
   u^{-\fft1{d-3}}\, (dr^2 + r^2\, d\Omega_{d-2}^2)\,,\nn\\
&=& -uf (dx^0)^2+ u^{-\fft1{d-3}}\, (f^{-1} -1)\, \fft{x_i\, x_j}{r^2}\, 
  dx^i\, dx^j + u^{-\fft1{d-3}}\, dx^i\, dx^i\,,\nn\\
&=& \eta_{\mu\nu}\, dx^\mu \, dx^\nu + h_{\mu\nu}\, dx^\mu\, dx^\nu\,,
\eea
with
\bea
h_{00}=1-u f\,,\qquad h_{ij}= (u^{-\fft1{d-3}}\, -1)\, \delta_{ij} +
  u^{-\fft1{d-3}}\, \, (f^{-1}-1)\, \fft{x_i\, x_j}{r^2}\,,\qquad
h_{0i}=0\,.\label{hbh}
\eea

  The metric functions $f$ and $u$ have the form
\bea
f= 1- m\, \rho\,,\qquad u= 1 -\sigma\,\rho + {\cal O}(\rho^2)\,,\qquad
\hbox{where}\qquad \rho=\fft1{r^{d-3}}\,,
\eea
and $m$ and $\sigma$ are constants.  If we assume that $r$ is very large,
so that the metric is nearly Minkowski, we can focus just on terms up to
linear order in $\rho$.  We then have
\bea
h_{00}= (m+\sigma)\, \rho +\cdots\,,\qquad h_{ij}= \fft{\sigma}{d-3}\, 
\delta_{ij}\, \rho + m\,\rho\, \fft{x_i\, x_j}{r^2} + \cdots\,,
\eea
which implies 
\bea
\del_j h_{ij}= (m-\sigma)\, \fft{x_i}{r^{d-1}} +\cdots\,.
\eea
Applying the ADM mass formula we find
\bea
M_{ADM} &=& \fft1{16\pi}\, \int_\Sigma 
  d\Sigma_i\, (\del_j h_{ij} - \del_i h_{jj})
\,,\nn\\
&=& \fft1{16\pi}\, \int_\Sigma (d-2)(m+\sigma)\, d\Omega_{d-2}\,,\nn\\
&=& (d-2)(m+\sigma)\, \fft{\Omega_{d-2}}{16\pi}\,.
\eea
Applied to the $SL(3,R)$ Toda black holes this gives
\bea
M_{ADM}= \fft{(d-2)\,\Omega_{d-2}}{16\pi}\, 
 \fft{(1-\beta_1)(1-\beta_2)(1-\beta_1\beta_2)}{\gamma_1\gamma_2}\,,
\eea
in agreement with the expression given in \cite{luyang}.

  In \cite{Heidenreich:2020upe}, 
the black hole mass is calculated using the
fact that in De Donder gauge, the trace-reversed lineralised metric
perturbation $\bar h_{\mu\nu}\equiv h_{\mu\nu}-\ft12 h\, \eta_{\mu\nu}$,
satisfying the gauge condition $\del^\mu\bar h_{\mu\nu}=0$, 
 obeys
$\square \bar h_{\mu\nu} = -16\pi\, T_{\mu\nu}$.  
In particular, in \cite{Heidenreich:2020upe} 
the mass $M_H$ is read off from the expression
%
%
(setting $\kappa^2=\ft12$ to accord with our 
conventions)
\bea
\bar h_{00} = \fft{M_H}{(d-3)\, \Omega_{d-2}}\, \fft1{r^{d-3}}
 +\cdots \,.\label{Hmass}
\eea

The linearised metric 
$h_{\mu\nu}$ given in eqns (\ref{hbh}) is in fact not in De Donder 
gauge: We have
\bea
h= -h_{00} + h_{ii} = \fft{2\sigma}{d-3}\, \rho +\cdots \,,
\eea
and since the metric is static we need only check
\bea
\del_j \bar h_{ij} =\del_j h_{ij} -\del_i h= \fft{m\, x_i}{r^{d-1}} +\cdots\,,
\eea
which is non-zero at this leading order.  We can easily make a coordinate 
transformation to put $h_{\mu\nu}$ in De Donder gauge at the leading order,
by considering
\bea
h_{\mu\nu} \longrightarrow h_{\mu\nu}'= h_{\mu\nu} + \del_{(\mu}\xi_{\nu)}\,,
\eea
Let us try taking $\xi_\mu$ to be given by 
\bea
\xi_0=0\,,\qquad \xi_i= \fft{\alpha\, x_i}{r^{d-1}}\,,
\eea
where $\alpha$ is a constant to be determined.  Thus we shall have
\bea
h\longrightarrow h'= h + \fft{2\alpha}{r^{d-3}}\,,
\eea
and so
\bea
\del_j \bar h_{ij}\longrightarrow \del_j \bar h'_{ij}=
  \del_j \bar h_{ij} -\fft{\alpha\, x_i}{r^{d-1}} =
  \fft{[m-(d-3)\, \alpha]\, x_i}{r^{d-1}}+\cdots \,,
\eea
implying that $\bar h_{\mu\nu}'$ will be in De Donder gauge at leading order
if we choose $\alpha = m\, (d-3)^{-1}$.  Finally, note that this gives
\bea
\bar h_{00}' = \bar h_{00} + \fft{m}{d-3}\, \rho +\cdots\,,
\eea
and so
\bea
\bar h_{00}' &=& \Big[ m +\fft{(d-2)\,\sigma}{d-3}\Big] \rho 
            + \fft{m}{d-3}\, \rho +\cdots\nn\nn\\
&=& \fft{(d-2)\,(m+\sigma)}{d-3}\,\rho+\cdots\,.
\eea
Thus, we find that the mass $M_H$ in
\cite{Heidenreich:2020upe} is given by
\bea
M_H= (d-2)\, \Omega_{d-2}\, (m+\sigma) = 16\pi\, M_{ADM}\,.
\eea
  In terms of the rescaled mass $M$ that we defined in eqn (\ref{rescaled}),
we therefore have
\bea
M_H= \ft12 (d-2)\, \Omega_{d-2}\,M\,.
\eea

\subsection{Scalar Charge}

  The role of scalar charge in black hole thermodynamics was described in
\cite{gibkalkol}.  
  In eqn (4.16) of \cite{Heidenreich:2020upe}, the scalar charge
$\mu= M'(\phi_0)$ is read off from the leading falloff term in the
large-distance expansion of the scalar field:
\bea
\phi=\phi_0 -\fft{G_{\phi\phi}^{-1}\, M'(\phi_0)}{(d-3)\, \Omega_{d-2}}\,
  \fft1{r^{d-3}} +\cdots\,.
\eea
If we write the large-distance forms of the two functions $H_1$ and $H_2$
in the solution (\ref{sl3rsol}) as 
\bea
H_i = 1 + \ell_i\, \rho +\cdots\,,
\eea
then we see from the expression for $\phi$ in eqns (\ref{sl3rsol}) that
\bea
\phi = \ft12 \sqrt{\fft{3(d-2)}{2(d-3)}}\, (\ell_2-\ell_1)\,\rho+\cdots\,,
\eea
and therefore the scalar charge is
\bea
\mu = M'(\phi_0) = \sqrt{\fft{3(d-2)(d-3)}{8}}\, \Omega_{d-2}\, 
(\ell_2-\ell_1)\,.
\eea
Now, for the solution (\ref{sl3rsol}) we have
\bea
\ell_1 = \fft{2\beta_1\, (1-\beta_2)\, m}{\gamma_1}\,,\qquad
\ell_2 = \fft{2\beta_2\, (1-\beta_1)\, m}{\gamma_2}\,,
\eea
and so the scalar charge is given by
\bea
\mu= M'(\phi_0) = \sqrt{\fft{3(d-2)(d-3)}{2}}\, 
     \fft{(\beta_2-\beta_1)(1-\beta_1\,\beta_2)\, m}{\gamma_1\,\gamma_2}\,.
\eea
In terms of the rescaled scalar charge $\Sigma$ that we defined in
eqns (\ref{rescaled}), we therefore have
\bea
\mu= M'(\phi_0) = \sqrt{\fft{(d-2)(d-3)}{2}}\,\Omega_{d-2}\, \Sigma\,.
\eea

\subsection{Electric Charges}

   In eqn (4.14) of \cite{Heidenreich:2020upe}, the 
electric charge $Q_H$ is read off from 
the asymptotic form of the electromagnetic potential $A= -\Phi\, dt$,
with
\bea
\Phi = \Phi_0 + \fft{Q_H}{(d-3)\, \Omega_{d-2}}\, \fft1{r^{d-3}} +\cdots\,.
\eea
(We have set $e=1$ to accord with our conventions, and also we have 
allowed for a possible constant term $\Phi_0$ in the potential at infinity.)
Writing the gauge potentials $A_1$ and $A_2$ in our eqns (\ref{sl3rsol}) as
$A_i= \Phi_i\, dt$, we have
\bea
\Phi_1 &=& \hbox{const.} - \fft{[\beta_1\, m - (1-\beta_1)\,\ell_1]}{
\sqrt{\beta_1\,\gamma_2}}\, \sqrt{\fft{d-2}{d-3}}\, \rho +\cdots\nn\\
&=& \hbox{const.}  + \sqrt{\fft{d-2}{d-3}}\,
\fft{\sqrt{\beta_1\, \gamma_2}\, m}{\gamma_1} +\cdots\,,
\eea
with an equivalent expression for $\Phi_2$ obtained by exchanging the 1 and
2 subscripts.  Thus we have the charges $Q^H_i$ given by
\bea
Q^H_1 &=& \sqrt{(d-2)(d-3)}\,  \Omega_{d-2}\, 
\fft{\sqrt{\beta_1\,\gamma_2}\, m}{\gamma_1}\,,\nn\\
Q^H_2 &=& \sqrt{(d-2)(d-3)}\,  \Omega_{d-2}\,
\fft{\sqrt{\beta_2\,\gamma_1}\, m}{\gamma_2}\,.
\eea
In terms of the rescaled electric charges $Q_1$ and $Q_2$ 
defined in eqns (\ref{rescaled}), the charges are therefore given 
by
\bea
Q^H_1 &=& \sqrt{\fft{(d-2)(d-3)}{2}}\, \Omega_{d-2}\, Q_1\,,\nn\\
Q^H_2 &=& \sqrt{\fft{(d-2)(d-3)}{2}}\, \Omega_{d-2}\, Q_2\,.
\eea

\subsection{Force Between Widely-Separated Black Holes}

 From eqn (4.17) in \cite{Heidenreich:2020upe}, we will have, in our
$SL(3,R)$ Toda case,  $F_{12}^H = \cF_H/
r^{d-2}$, with
\bea
\cF_H = \fft1{\Omega_{d-2}}\,\Big[ Q^H_1\, \wtd Q^H_1 + Q^H_2\, \wtd Q^H_2
  -\mu\, \tilde\mu - \fft{(d-3)}{2(d-2)}\, M_H\, \wtd M_H\Big]\,.\label{heidF}
\eea
Expressed in terms of the rescaled masses and charges of our eqn 
(\ref{rescaled}), we therefore have
\bea
\cF_H = \ft12(d-2)(d-3)\, \Omega_{d-2}\, \cF\,,
\eea
with the force coefficient given by
\bea
\cF = Q_1\, \wtd Q_1 + Q_2\,\wtd Q_2 - \Sigma\, \wtd\Sigma 
-\ft14 M\, \wtd M\,,
\eea
as given earlier in eqn (\ref{cF}).

\section{BPS Black Holes}

  For comparison with the results for the forces between the non-BPS extremal
black holes that we have been considering in this paper, 
it is interesting to look at examples of
BPS extremal black holes.  A convenient and rather general class of examples
can be constructed as solutions for the following theories, which were 
discussed in \cite{lptx}.  The $d$-dimensional Lagrangian is
\bea
{\cal L} = R -\ft12 (\del\vec\phi)^2 - \ft14 \sum_{\alpha=1}^N 
e^{\vec c_\alpha\cdot\vec\phi}\, F_\alpha^2\,,\label{genlag}
\eea
where $\vec\phi$ denotes an $(N-1)$-vector of dilatonic scalar fields,
and the constant vectors $\vec c_\alpha$ describing the couplings of
these scalars to the $N$ 2-form gauge fields $F_\alpha$ obey the relation
\bea
\vec c_\alpha\cdot\vec c_\beta = 4\delta_{\alpha\beta} - \fft{2(d-3)}{d-2}\,.
\label{cdotc}
\eea
In dimensions $d\le 10$ these Lagrangians arise as consistent truncations
of the maximal supergravities obtained by the toroidal reductions of
$d=11$ supergravity.  One can also consider theories of the form 
(\ref{genlag}) more generally, in any arbitrary dimension.

The extremal $N$-charge black hole solutions of the theory (\ref{genlag}) 
are given by \cite{lptx}
\bea
ds^2 &=& -\Big(\prod_\alpha H_\alpha\Big)^{-\fft{d-3}{d-2}} \, dt^2
  + \Big(\prod_\alpha H_\alpha\Big)^{\fft1{d-2}}\, 
                 (dr^2 + r^2 d\Omega_{d-2}^2)\,,\nn\\
F_\alpha &=& dt\wedge dH_\alpha^{-1}\,,\qquad 
 \vec\phi= \ft12 \sum_\alpha \vec c_\alpha\, \log H_\alpha\,,
\eea
where
\bea
H_\alpha = 1 + \fft{q_\alpha}{r^{D-3}}\,.
\eea
Calculating the ADM mass as discussed in appendix A we find
\bea
M_{ADM} = \fft{(d-3)\, \Omega_{d-2}}{16\pi}\, \sum_\alpha q_\alpha\,.
\eea
The mass $M_H$, scalar charges $\vec \mu$ and electric
charges $Q^H_\alpha$, 
calculated as described in appendix A, are therefore given by
\bea
M_H &=& (d-3)\,\Omega_{d-2}\,  \sum_\alpha q_\alpha\,,\nn\\
\vec \mu &=& \ft12 (d-3)\,\Omega_{d-2}\, 
                 \sum_\alpha \vec c_\alpha\, q_\alpha\,,\nn\\
Q^H_\alpha &=& (d-3)\, \Omega_{d-2}\, q_\alpha\,.
\eea
The force between two distant BPS black holes, defined as in the 
formula (\ref{heidF}), gives
\bea
\cF_H = (d-3)^2\, \Omega_{d-2}\, \Big[ \sum_\alpha q_\alpha\,\tilde q_\beta 
- \ft14 \sum_{\alpha,\beta} \vec c_\alpha\cdot \vec c_\beta\, 
q_\alpha\, \tilde q_\beta-
\fft{d-3}{2(d-2)}\, \sum_{\alpha,\beta} q_\alpha \,\tilde q_\beta\Big]\,.
\eea
After making use of the relation (\ref{cdotc}), we see that the
force between any pair of BPS black holes is equal to zero.  This should be
contrasted with the situation for the $SL(3,R)$ extremal black holes,
which are not BPS, where the force is zero only if the electric charges of
one black hole are an overall multiple of the electric charges of 
the other black hole.

  It can easily be verified that more generally, one can consider 
configurations of the form
\bea
ds^2 &=& -\Big(\prod_\alpha H_\alpha\Big)^{-\fft{d-3}{d-2}} \, dt^2
  + \Big(\prod_\alpha H_\alpha\Big)^{\fft1{d-2}}\, dy^i\, dy^i\,,\nn\\
F_\alpha &=& dt\wedge dH_\alpha^{-1}\,,\qquad
 \vec\phi= \ft12 \sum_\alpha \vec c_\alpha\, \log H_\alpha\,,
\eea
where $y^i$ are coordinates on the Euclidean $(d-1)$-dimensional 
transverse space, and the functions $H_\alpha$ depend on the 
$y^i$ coordinates.  The equations of motion following from the Lagrangian
(\ref{genlag}) are satisfied if the $H_\alpha$ are arbitrary harmonic
functions on the $(d-1)$-dimensional Euclidean space.  Multi-black hole
solutions are obtained by taking the $H_\alpha$ to be sums of elementary
harmonic functions of the form 
\bea
H_\alpha = \sum_{a} \fft{q_{\alpha a}}{|\vec y -\vec y_{\alpha a}|^{d-3}}\,.
\eea
(Global considerations impose certain constraints on the strengths
$q_{\alpha  a}$ of the elementary functions located at the mass-centres
$\vec y_{\alpha a}$.)  It should be emphasised that the ability to
construct multi-centre solutions of this form is intimately related 
to the fact that the force between any pair of single-centre black holes
is zero.  This is very different from the situation for the non-BPS 
extremal black holes we have been considering in this paper.

\end{document}